\documentclass[12pt]{article}

\usepackage{amsmath}
\usepackage{amsfonts}
\usepackage{url}
\usepackage{circuitikz}
\usepackage{endnotes}
\usepackage{framed}
\usepackage{graphicx}
\usepackage{caption}
\usepackage[version=4]{mhchem}

\usepackage{titlesec}
\titleformat{\section}{\normalsize\bfseries}
   {\thesection}{1em}{}
\titleformat{\subsection}{\normalsize\bfseries}
   {\thesubsection}{1em}{}

\begin{document}

\setlength{\parindent}{0in}
\setlength{\parskip}{0.4in plus0.2in minus0.2in}
\setlength{\voffset}{0in}
\setlength{\topmargin}{0in}
\setlength{\headheight}{0in}
\setlength{\headsep}{0in}
\setlength{\footskip}{0.5in}

\noindent

\bibliographystyle{plain}


\long\def\symbolfootnote[#1]#2{\begingroup%
\def\thefootnote{\fnsymbol{footnote}}
     \footnote[#1]{#2}\endgroup}




\begin{center}
{\large\textbf{On Calculating the Current-Voltage Characteristic\\
of Multi-Diode Models for Organic Solar Cells}}\\
---------------------------------------------------------------------------------------------\\
Ken Roberts\footnote{Physics and Astronomy Department, 
Western University, London, Canada, krobe8@uwo.ca},
S. R. Valluri\footnote{Physics and Astronomy Department,
Western University, London, Canada;
King's University College, London, Canada, 
valluri@uwo.ca, vallurisr@gmail.com}\\
December 11, 2015
\end{center}

\begin{abstract}


We provide an alternative formulation of the
exact calculation of the current-voltage characteristic 
of solar cells which have been modeled with
a lumped parameters equivalent circuit 
with one or two diodes.
Such models, for instance, are suitable for describing organic
solar cells whose current-voltage characteristic curve
has an inflection point, also known as an S-shaped anomaly.
Our formulation avoids the risk of numerical overflow
in the calculation.
It is suitable for implementation in Fortran, C or on micro-controllers.
\end{abstract}


\section{Introduction}
\label{sect-intro}

The current-voltage characteristic of a solar cell is often
modeled using an equivalent circuit with lumped parameters.
Different models use one, two or three diodes in the circuit.
These models have an associated implicit equation which
relates the current and voltage measurements, and 
parameters which can be adjusted to make the model
fit experimental data.
The implicit equations for some models can be solved 
explicitly to obtain an exact equation $V = f(I)$ by which the 
voltage $V$ can be calculated from the current $I$.

Some of the formulas for the equation $V = f(I)$, as they are
usually written, involve intermediate calculations which,
if not handled properly, may produce arithmetic overflow.
Overflow may happen even if the calculation is done using
double or quadruple precision hardware floating point
arithmetic.
It may be necessary to use special software, 
such as a symbolic mathematics package or a subroutine
library for multi-precision arithmetic, in order to calculate
some of the formulas for $V = f(I)$ as usually written.
That can restrict the applicability of the explicit formula
to situations which have appropriate computer software
resources.

The purpose of this paper is to review some of the formulas
for $V = f(I)$, and to rewrite them to avoid possible overflow.
The rewritten formulas can be utilized in Fortran or C with
hardware floating point,
or on micro-processors with fixed point arithmetic, 
in order to calculate $V = f(I)$ with little risk of overflow.
That enables the explicit formulas to be used in field
implementations, such as in test equipment for solar cells,
or for load balancing of solar energy installations.
There is no need to have a Lambert W function
implementation in the Fortran, C, or micro-processor
environment.
We give a simple algorithm for calculation of an analytic
function $y = g(x)$ which serves the same purpose.

We will start with the one diode model in section 
\ref{sect-1diode}.
That model is often used for non-organic solar 
cells.
We show how one can rewrite 
a $V = f(I)$ formula to avoid overflow.
Rewriting of the one diode
$V = f(I)$ formula for computational robustness 
lays the groundwork for 
considering the two diode model situation.

In section \ref{sect-2diode} 
we will turn to a particular two diode model.
For some organic solar cells the experimentally measured
$I-V$ characteristic curve may have an inflection point, 
also called an S-shaped anomaly.
A two diode model can be used to represent the
shape of the $I-V$ curve and to achieve a good fit of the
model to the observational data.
An inflection point is frequently associated with poor
performance of the organic solar cell.  
The inflection point can be altered, or even removed,
by annealing of the solar cell.
Having a good analytic formula $V = f(I)$ for the $I-V$
characteristic may be helpful for understanding how the
annealing process works.  
As well, because two diode models have
more parameters than the one diode model, 
it may not be as easy to use numerical search techniques
to identify the
relevant subspace of physically achievable solar cells.

Our primary objective in this paper is to obtain methods
that will assist with the computational tasks associated
with investigation of organic solar cells, in particular
using two diode models.
We also expect that our methods will be useful for field
implementations of load balancing for solar panel arrays
of solar cells, using either one or two diode models.

\section{One Diode Model}
\label{sect-1diode}

\begin{figure}
\makebox[\textwidth][c]{
  \includegraphics[scale=1.00]{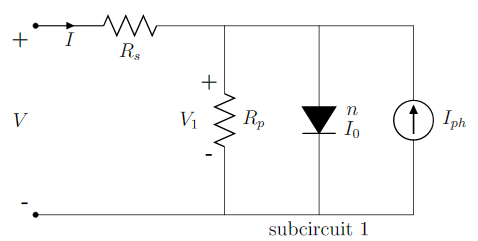}}
{
\captionsetup{format=hang}
  \caption{
\label{fig-ckt-1diode}One diode model of a solar cell.\\
Diode reverse saturation current $I_0$ and ideality factor $n$.
\\
  --------------------------------------------------------------------------------------
}}
\end{figure}

Solar cells can be modeled via an equivalent circuit
with lumped parameters and a single diode
\cite{Nelson-2004}, as shown in figure \ref{fig-ckt-1diode}.
This model is described by an implicit equation which
relates the current $I$ and the voltage $V$ in terms
of the cell parameters:
\begin{equation}
\label{eq1diode}
  I = I_0\,\Big(
           \mathrm{exp}\Big[\frac{q}{n k_B T}(V - IR_s)\Big] -1\Big)
       + \frac{V - IR_s}{R_{sh}} - I_{ph},
\end{equation}
where $I_{ph}$ is the cell's photocurrent,
$I_0$ is the diode's reverse saturation current,
$n$ is the diode's ideality factor,
$R_s$ is the series resistance,
$R_p$ is the parallel (shunt) resistance,
$q$ is the magnitude of the electron charge,
$k_B$ is Boltzmann's constant,
and $T$ is absolute temperature.
Here we have written the current $I$ with the opposite
sign to the current convention on pp 14-15 of \cite{Nelson-2004}.
As well, we have assumed unit solar cell area, so that
we speak of current rather than current areal density.
This model is descriptive of the solar cell under some
assumed standard illumination, appropriate for the
experimental tests which provide the $I,V$ data points.
The photocurrent generated within the solar cell under
illumination is represented by $I_{ph}$.

This one diode model is adequate for the description 
of many solar cells.
The implicit equation (\ref{eq1diode}) can be solved to obtain 
an explicit equation for $I$ as a function of $V$, or an
explicit equation for $V$ as a function of $I$
\cite{JK-2004}.
Here we will focus on just the explicit equation
for $V$ as a function of $I$, as that is the form which is
used in two diode models.

Jain and Kapoor \cite{JK-2004} give an exact explicit
formula for $V$ as a function of $I$.
In our notation and with our sign convention,
their formula is
\begin{eqnarray}
  \nonumber
  V = f(I) &=& IR_s + (I+I_{ph}+I_0)R_p\\
    \label{eqjkvf}
         &-& \frac{n k_B T}{q}
          W\Big(\frac{q}{n k_B T} I_0 R_p
                     \mathrm{exp}\Big[
                         \frac{q}{n k_B T} R_p (I + I_{ph} + I_0)
                                         \Big]
             \Big)
\end{eqnarray}
The function $W$ is the Lambert W function
\cite{Corless-1996, Valluri-2000, NIST-2010}.
The principal branch of that function is 
used in these calculations since the argument
to $W$ is a positive real value and the result is
also to be a real value.

\begin{figure}
\makebox[\textwidth][c]{
  \includegraphics[scale=0.60]{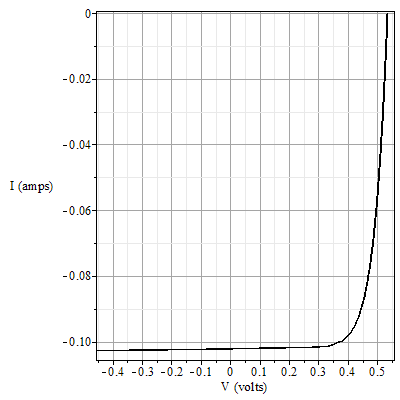}}
{
\captionsetup{format=hang}
  \caption{
\label{fig-jk1-blue}I-V characteristic for the one diode model
of the ``blue" solar\\
cell of \cite{CAMM-1981}, 
calculated from the fitted parameters described in \cite{PCP-1984}.
\\
  -------------------------------------------------------------------------------------
}}
\end{figure}

Suppose we consider a solar cell with parameters 
$I_{ph} = 0.1023$ amp, 
$I_0 = 0.1036 \times 10^{-6}$ amp,
$n = 1.5019$, 
$R_s = 0.06826$ ohm,
$R_p = 1000$ ohms,
at $T = 300$ K.
These values are the parameters obtained in \cite{PCP-1984}
for the experimental data for the ``blue" solar cell.
The original experimental data, and the previous model fit
via solution of an equivalent of the implicit equation 
(\ref{eq1diode}), are reported in \cite{CAMM-1981}.
The $I-V$ characteristic curve for this fitted model
with the above parameters from reference \cite{PCP-1984}
is shown in figure \ref{fig-jk1-blue}.

\begin{figure}
\makebox[\textwidth][c]{
  \includegraphics[scale=0.50]{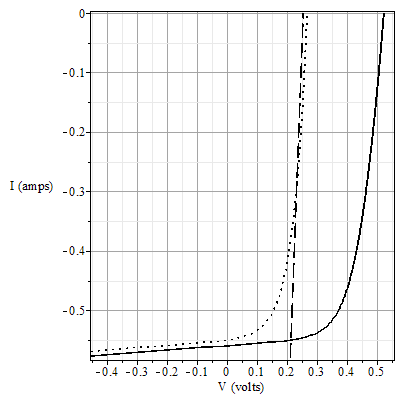}}
{
\captionsetup{format=hang}
  \caption{
\label{fig-jk1-grey-sum}I-V characteristic for the one diode model
of the ``grey" solar\\
cell of \cite{CAMM-1981}, 
calculated from the fitted parameters described in \cite{PCP-1984}.\\
Shows the two components of the model; a sloped straight line\\
and a ``J"-shape representing the scaled $g(x)$ function.\\
The sum of these two components is the grey cell's $I-V$ curve.
\\
  -------------------------------------------------------------------------------------
}}
\end{figure}

As a second example, we
consider another solar cell, the ``grey" solar cell,
which was also experimentally studied and fitted, 
with results reported in \cite{CAMM-1981} and further
considered in \cite{PCP-1984}.  
The parameters obtained in \cite{PCP-1984} for the
grey solar cell are 
$I_{ph} = 0.5610$ amp, 
$I_0 = 5.514 \times 10^{-6}$ amp,
$n = 1.7225$, 
$R_s = 0.07769$ ohm,
$R_p = 25.9$ ohms,
at $T = 307$ K.
The $I-V$ characteristic of the fitted model of the grey solar cell
is shown as the solid curve in figure \ref{fig-jk1-grey-sum}.

If the $I-V$ characteristic curve for the blue solar cell,
graphed in figure \ref{fig-jk1-blue}, 
were to be calculated using the formula given in 
equation (\ref{eqjkvf})
using IEEE-754-2008 standard \cite{IEEE-754} 
hardware floating point in double precision, 
there would be an arithmetic overflow exception.
For a current $I = 0$, the argument of the $W()$ function is
$4.59 \times 10^{1141}$.
The maximum representable magnitude in IEEE-754-2008
compliant double precision is about $10^{323}$.
One might use quadruple precision floating point,
referred to in the standard as ``binary128", as it is able
to represent magnitudes up to about $10^{5107}$.
However, as we will see with another example below,
also using model parameters which were obtained via
fitting the two diode model to experimental data from 
an actual solar cell, 
even quadruple floating point arithmetic can be insufficient
to avoid overflow during the calculation of the $V = f(I)$ function
via a formula such as that written in equation (\ref{eqjkvf}).

On the other hand, the calculation of the grey solar cell's
characteristic curve, shown in figure \ref{fig-jk1-grey-sum},
would not produce an arithmetic overflow.  
The maximum magnitude involved in the calculation
of that curve via formula (\ref{eqjkvf}) is about $3 \times 10^{138}$,
even for intermediate variables involved in the calculation.
That is compatible with the limitations of standard-compliant
hardware double precision arithmetic.

How is one to know when it is safe to use formula (\ref{eqjkvf})
for a calculation?  The blue solar cell is not an anomaly.
In fact, the authors of the original study \cite{CAMM-1981}
consider the blue cell to be of better quality compared
to the grey cell.
The blue cell has a larger fill factor, lower series
resistance, higher shunt resistance, and lower ideality factor.

Higher shunt resistance in the lumped parameters model is
associated with improvements in solar cell quality.
However, higher shunt resistance also increases the
numerical magnitude of the Lambert W function argument
in formula (\ref{eqjkvf}).
Solar cells are better now, three decades after
the study \cite{CAMM-1981}.
Thus one can expect to encounter actual solar cells whose
model parameters will produce arithmetic overflow if the
formula (\ref{eqjkvf}) is applied in a standard computer
language such as Fortran or C, or if an implementation of
that formula is attempted on a micro-controller.

How did the authors of the studies \cite{CAMM-1981} and
\cite{PCP-1984} utilize the model parameters they derived?
They did not use the explicit formula (\ref{eqjkvf}), instead relying
upon solving the implicit equation (\ref{eq1diode}), and
other approximation techniques.  
One would like, however, to be able to use a explicit 
and exact formula such as (\ref{eqjkvf}) because
it is analytic, that is has derivatives of all orders, and hence
can be used in optimization studies and other investigations.
What is desired is to obtain a formula to replace (\ref{eqjkvf})
which is exact, explicit and analytic, yet is also robust when
used in calculations, and does not pose a risk of arithmetic
overflow.   Such a formula is desirable both for laboratory
studies, and for industrial applications (test equipment for
a manufacturing line) and field installations (load balance).
It should be suitable for programming in Fortran or C, or
on micro-controllers which may have only a fixed point
arithmetic software package.

A related concern is possible cancellation causing loss of 
significant digits in the result.
Equation (\ref{eqjkvf}) involves some subtractions.
For $I = 0$ for the blue solar cell, we have a value
of 102.300 for the linear terms, from which is subtracted
a value 101.764 representing the scaled value of the
Lambert W function, causing a loss of about two
significant digits.  
Cancellation is not problematic in this instance.
However, if there were a way to reduce cancellation
while rewriting the formula (\ref{eqjkvf}), that would be desirable.

Examples such as these have motivated our introduction
of an alternative technique for calculating a function
like $V = f(I)$ as defined by equation (\ref{eqjkvf}).
In the next section, 
we use the function $g(x) = \mathrm{log}(W(e^x))$
which was described in a previous note \cite{Roberts-2015},
and rewrite the formula $V = f(I)$ given by
(\ref{eqjkvf}) in terms of the function $g(x)$.
Here $\mathrm{log}$ denotes the natural logarithm
(base $e$) function, the inverse of the exponential function.

\section{Rewrite of Equation (\ref{eqjkvf}) Formula}
\label{sect-rewrite-jk}

First, we make the general observation that the function
$g(x)$ satisfies
\begin{equation}
\label{eqgx}
  g(x) = \mathrm{log}(W(e^x)) = x - W(e^x).
\end{equation}
This fact is obtained in \cite{GSOCM-2006} as the
equivalence of equations (32) and (37) of that paper.
To verify equation (\ref{eqgx}) directly, it suffices to take the exponential
of both sides of (\ref{eqgx}) , multiply out,
and use the defining property of the Lambert W function,
that $W(z)\,e^{W(z)} = z$.
Hence, when we have a formula containing 
an expression of the form
$x - W(e^x)$ 
we can replace it by calculation of the function $g(x)$.
The two forms are mathematically equivalent, but
the form $x - W(e^x)$ has a risk of arithmetic overflow
when the argument $e^x$ to the $W()$ function
is calculated.
There is also a risk of possible cancellation causing loss of
significant digits when the subtraction is performed.
The function $g(x)$
is well-behaved in numerical calculations.

The function $y = g(x)$ is just the principal branch of
the real values of the Lambert W function in another
coordinate system.
We have $e^y = W(e^x)$.
Thus if the function $w = W(z)$, with $w$ and $z$ 
restricted to positive real values, were to be graphed
on log-log axes, we would see the graph of the
$y = g(x)$ function.
The distinction between the two functions is not in their
mathematical properties, but in their computational
practicality.
In section \ref{sect-galgo} below 
we describe how to compute $g(x)$
without overflow.
Further details and discussion of
the function $g(x)$ are given in 
reference \cite{Roberts-2015}.

We wish to recast (\ref{eqjkvf}) to evaluate $W(e^{x})$
for some $x$.
Clearly we should write
\begin{equation}
\label{eqx}
  x = \mathrm{log}\Big[\frac{q}{n k_B T}\,I_0 R_p
                \Big]
       + \frac{q}{n k_B T}\,R_p\,(I + I_{ph} + I_0).
\end{equation}
Transposing the $I R_s$ term in equation (\ref{eqjkvf}) 
and multiplying by
$\frac{q}{n k_B T}$ we obtain
\begin{eqnarray*}
  \frac{q}{n k_B T}\,(V - I R_s)
        &=& \frac{q}{n k_B T}\,(I + I_{ph} + I_0) \, R_p
                - W(e^{x})\\
        &=& \Big(x
                        - \mathrm{log}\Big[
                              \frac{q}{n k_B T}\,I_0 R_p
                           \Big]\Big)
                  - W(e^{x})\\
        &=& g(x) - \mathrm{log}\Big[
                              \frac{q}{n k_B T}\,I_0 R_p
                           \Big].
\end{eqnarray*}
That is,
\begin{equation}
\label{eqv}
  \frac{q}{n k_B T}\,(V - I R_s)
         + \mathrm{log}\Big[
                              \frac{q}{n k_B T}\,I_0 R_p
                           \Big] = g(x).
\end{equation}

Consider the geometric content of equations 
(\ref{eqx}) and (\ref{eqv}).
Equation (\ref{eqx}) says that the value of the variable $x$
is obtained by a linear shift and scale change of the
current $I$.
Equation (\ref{eqv}) says that the value of the function $g(x)$
is obtained by a linear shift and scale change of the
subcircuit 1 voltage drop $V_1 = V - IR_s$.
Hence we expect the graph of $V_1 = V - IR_s = f(I) - IR_s$ to
look like the graph of $y = g(x)$, but with a shift of
origin and with linear scale changes on the two axes.

\begin{figure}
\makebox[\textwidth][c]{
  \includegraphics[scale=0.60]{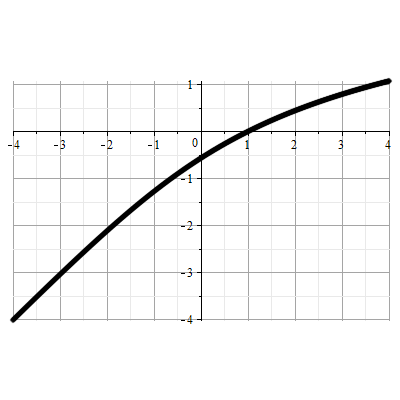}}
{
\captionsetup{format=hang}
  \caption{
\label{fig-logwexp1}Graph of 
  $y = g(x) = \mathrm{log}(W(\mathrm{exp}(x)))$\\
  for moderate magnitudes of the argument.\\
  --------------------------------------------------------------------------------
}}
\end{figure}

\begin{figure}
\makebox[\textwidth][c]{
  \includegraphics[scale=0.60]{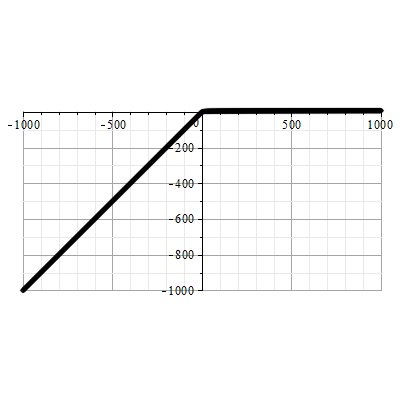}}
{
\captionsetup{format=hang}
  \caption{
\label{fig-logwexp2}Graph of 
  $y = g(x) = \mathrm{log}(W(\mathrm{exp}(x)))$\\
  for large magnitudes of the argument.\\
  --------------------------------------------------------------------------------
}}
\end{figure}

Figures \ref{fig-logwexp1} and \ref{fig-logwexp2}
show the graph of $y = g(x)$ at
two different magnifications.
From close up, $-4 \le x \le 4$ in figure \ref{fig-logwexp1}, 
the graph looks like a smooth curve going 
through the point $(1,0)$ near the origin.
From far away, $-1000 \le x \le 1000$ in figure \ref{fig-logwexp2},
the graph looks like a pair of straight lines, represented
by an abrupt change in slope at the origin.
Which picture we use for matching with a part of the 
$V - IR_s = f(x) - IR_s$ curve, 
that is either a smooth curve
or a pair of straight lines joined at what appears to
be a sharp corner,
depends upon the particular scales used for $I$ and $V$.
That behavior is of course modified by the physical
limitations on the voltages and currents within the device
being modeled.
However, regardless of appearance, whether a smooth curve
or joined straight lines, the function $g(x)$ is analytically
smooth.
That is, $g(x)$ has an unlimited number of derivatives,
and one can use it to determine
extrema or inflection points, and so on.

The voltage $V$ across the solar cell is given
in terms of the function $g(x)$, using the 
variable $x$ which is a
transformation of the current $I$, 
and is calculated as
\begin{eqnarray}
  \nonumber
  V = I\,R_S &+& n \, \frac{k_B T}{q}\,g(x)\\
                  \label{eqv1d}
                  &-&  \frac{n k_B T}{q} \mathrm{log}\Big[
                              \frac{q}{n k_B T}\,I_{0}R_{p}
                          \Big].
\end{eqnarray}

Equation (\ref{eqv1d}), with the ancillary equation
(\ref{eqx}) to define $x$,
is our proposed replacement for the formula
(\ref{eqjkvf}).
The function $g(x)$ is to be implemented as
described in section \ref{sect-galgo}.
The two formulas (\ref{eqjkvf}) and (\ref{eqv1d})
are mathematically equivalent, 
and each of them explicitly calculates $V = f(I)$, the exact
solution for the one diode model.
The formulas differ, however, in that our proposed formula
is unlikely to experience overflow, and will likely have less
cancellation error.

When graphing the blue solar cell's $I-V$ curve
to obtain figure \ref{fig-jk1-blue}, 
formula (\ref{eqjkvf}) can have intermediate values
as large as $1.7 \times 10^{1144}$ in magnitude, which will
produce arithmetic overflow in standard-compliant
hardware double precision arithmetic.
In contrast, the calculation of this graph using
formulas (\ref{eqx}) and (\ref{eqv1d}) involves
intermediate values only up to 3000 in magnitude.
For the grey solar cell, the calculation of the graph
in figure \ref{fig-jk1-grey-sum} using the rewritten
formulas involves intermediate
values only up to 400 in magnitude,
whereas the original formula involved intermediate
values more than $10^{138}$ in magnitude.
These low magnitudes of intermediate variables
in calculating formulas (\ref{eqx}) and (\ref{eqv1d})
hold for the calculations within both
the portion of the code which implements the
evaluation of the $g(x)$ function itself,
and the portion of the code which carries out the
rest of the work in formulas (\ref{eqx}) and (\ref{eqv1d}).
The magnitudes in the latter portion of the
code can of course be adjusted
by a suitable choice of units.  The $g(x)$ function is in
essence a black box, just as the Lambert W function was
a black box in formula (\ref{eqjkvf}).
It is important, when one is going to use a black box
function, that it be reliable.

The moderate magnitudes of the intermediate values 
involved in computing
the $V = f(I)$ function using the $g(x)$ function
indicates the rewritten formula's suitability for implementation
in fixed point arithmetic, such as on a micro-controller. 

We can visualize formula (\ref{eqv1d}) geometrically
as the sum of two curves, a sloped straight line plus a copy
of the $y = g(x)$ curve.
First, recognize that we are graphing $V = f(I)$ with
the $V$ axis horizontal and the $I$ axis vertical, so the
copy of the $y = g(x)$ curve should be flipped over
the diagonal line $y = x$, to put $y$ horizontal and
$x$ vertical.
The $IR_s$ first term in (\ref{eqv1d}), with axes flipped,
is a straight line with slope $\frac{1}{R_s}$
through the origin, and becomes steeper if the
series resistance is lower.
The third term in (\ref{eqv1d}), the $\mathrm{log}()$ term,
is simply a shift of that straight line left or right (or up or down).
The second term of (\ref{eqv1d}) is the addition of a
copy of the flipped $y = g(x)$ curve 
proportional to the ideality factor $n$.
Addition of a flipped copy of the $y = g(x)$ curve 
provides the J-shape of the $I-V$ curve.

This discussion of the method of rewriting the one diode model 
lays the groundwork for rewriting of the formulas used for
two diode models.
We will see that similar techniques are applicable.
The matching of $I-V$ characteristics with one or two
inflection points can be interpreted as finding the 
appropriate scaling and shift parameters for adding
or subtracting flipped copies of the $g(x)$ function graph to a 
sloped straight line representing the series resistance.

Methodological Note:
The reader may wish to know how the above-mentioned 
intermediate value magnitudes in calculations were obtained.
We implemented the formulas (\ref{eqjkvf}),
(\ref{eqx}) and (\ref{eqv1d}) in a software package which
does not have a limitation on numerical magnitudes.
We inserted ``probe" code at various points in the
calculations, in order to record the magnitudes of
the intermediate values of interest.
These probe magnitudes were fed through a high-water-mark
filter, and recorded in global variables.
After a set of calculations, such as the preparation of a graph
like figures \ref{fig-jk1-blue} or \ref{fig-jk1-grey-sum},
the global variables were inspected to determine the
peak magnitudes which were involved in the calculations.
That enabled us to determine whether a hardware implementation
of the calculation would have produced arithmetic overflow.

\section{Two Diode Model}
\label{sect-2diode}

The current-voltage characteristic of some organic solar cells
shows an inflection point, also known as an ``S-shape anomaly".  
This has been modeled by a lumped parameters equivalent
circuit using two or three diodes.
Each model's equivalent circuit has an implicit equation relating
the current $I$ and voltage $V$.
Some of these model configurations are amenable to obtaining
exact and explicit formulas of the form $V = f(I)$,
utilizing the Lambert W function.
In this section we will consider a particular two diode
model, and discuss the rewriting of its formulas to reduce
the risk of arithmetic overflow during calculations.
Having an explicit function is useful because it enables
rapid and exact calculation of the $I-V$ curve given values
of the parameters of the model.  
Further the explicit function is analytic so it can be used
to perform optimization investigations.  There may be eight
(or more) parameters in a multi-diode model, 
so numerical random search 
optimization methods may be time consuming in comparison
to analytic methods.

\begin{figure}
\makebox[\textwidth][c]{
  \includegraphics[scale=1.00]{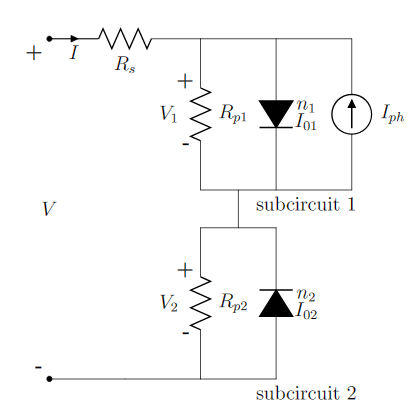}}
{
\captionsetup{format=hang}
  \caption{
\label{fig-ckt-2diode}Two diode model of a solar cell,
as in references \cite{RDPA-2012,GSLMMOC-2013}.\\
Subcircuit 1 has the photocurrent source, and a diode\\
with reverse saturation current $I_{01}$ and ideality factor $n_1$.\\
Subcircuit 2 has a reverse bias diode and a shunt resistance.
\\
  --------------------------------------------------------------------------------------
}}
\end{figure}

Romero, et al \cite{RDPA-2012} and
Garc\'ia-S\'anchez, et al \cite{GSLMMOC-2013}
have considered a two diode model 
with a reverse-bias second diode and shunt
resistor, as shown in figure \ref{fig-ckt-2diode}.
They give an exact solution utilizing eight parameters,
which we will write as
\begin{eqnarray}
  \nonumber
  V &=& f(I) = (I + I_{ph} + I_{01}) \, R_{p1} \\
        \nonumber
        &-& \frac{n_1 k_B T}{q} \,
          W\Big\{\frac{q}{n_1 k_B T}\,I_{01}R_{p1} \,
               \mathrm{exp}\Big[
                  \frac{q}{n_1 k_B T}\,R_{p1}\,(I + I_{ph} + I_{01})
                    \big]
             \Big\} \\
        \nonumber
        &+& \frac{n_2 k_B T}{q} \,
          W\Big\{\frac{q}{n_2 k_B T}\,I_{02}R_{p2} \,
               \mathrm{exp}\Big[
                  \frac{-q}{n_2 k_B T}\,R_{p2}\,(I - I_{02})
                    \Big]
             \Big\} \\
        \label{eqrdpa}
        &+& (I - I_{02}) \, R_{p2} + IR_s
\end{eqnarray}

We refer to the paper \cite{RDPA-2012} for the derivation.
All eight model parameters
$I_{ph}, R_s, I_{01}, R_{p1}, n_1, I_{02}, R_{p2}, n_2$
are positive numbers,
except for $R_s$ which can be positive or zero.

\begin{figure}
\makebox[\textwidth][c]{
  \includegraphics[scale=0.60]{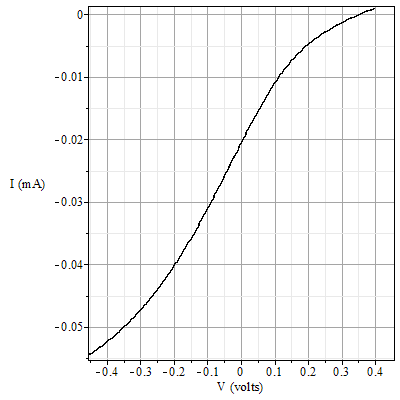}}
{
\captionsetup{format=hang}
  \caption{
\label{fig-rdpa2}Graph of $V = f(I)$ for two diode model
  of organic solar cell.\\
  The axes are switched; that is, $I$ is vertical and $V$ is horizontal.\\
  Agrees with best-fit curve shown in figure 2 
       of reference \cite{RDPA-2012}.\\
  --------------------------------------------------------------------------------
}}
\end{figure}

An example of the S-shape anomaly (inflection point) 
can be seen in figure
\ref{fig-rdpa2}, which has been calculated from the
model parameters obtained by Romero, et al \cite{RDPA-2012}
for a particular actual organic solar cell 
whose test data was fitted with their model.
The parameter values used for that figure are
$I_{ph} = 4.85\,\times\,10^{-5}$ amp (photocurrent);
$R_s = 0$ ohm (series resistance assumed equal to zero);
$I_{01} = 1.5\,\times\,10^{-5}$ amp
(primary diode reverse saturation current);
$R_{p1} = 1.0\,\times\,10^8$ ohms
(primary diode shunt resistance);
$n_1 = 2.4$ (primary diode ideality factor);
$I_{02} = 2.4\,\times\,10^{-7}$ amp
(secondary diode reverse saturation current);
$R_{p2} = 4.6\,\times\,10^4$ ohms
(secondary diode shunt resistance);
$n_2 = 9.5$ (secondary diode ideality factor).
The $I$ values range between about $-5.5\,\times\,10^{-5}$ amp
and $0.1\,\times\,10^{-5}$ amp,
and the corresponding $V$ values range between
about -0.42 volt and 0.42 volt.

The presence of an S-shape anomaly has been observed in organic
solar cells produced by a variety of methods, and is associated
with poor performance of the device.
See \cite{LMMNK-2010}
for background and for numerous references to examples
of organic solar cells manufactured by various methods.
The authors of \cite{LMMNK-2010} also explore the effect
of annealing on changes in the inflection point of a solar cell.
It is desirable to investigate the effect of annealing
on changes in the model parameters of an organic solar cell,
using the various two and three diode models available.
It should be mentioned that S-shape anomalies can also be seen
in non-organic solar cells. 
For instance, a recent paper \cite{KS-2015} on a silicon
quantum dot solar cell shows, in its figure 3, 
an inflection point in the $I-V$ characteristic of
the test device.

As with the one diode model, there can be arithmetic overflow
when the formula (\ref{eqrdpa}) is used for calculations.
Hence it is desirable to rewrite the formula using the $g(x)$ function.

The primary diode circuit
 (subcircuit 1 in figure \ref{fig-ckt-2diode})
is given by an equation identical to that used for a one diode
model, with $V_1$ denoting the voltage drop across that
subcircuit.  That is (formula (8) of \cite{RDPA-2012}),
\begin{eqnarray}
  \nonumber
  V_1 &=& f_1(I) = (I + I_{ph} + I_{01}) \, R_{p1} \\
        \label{eqrdpa8}
        &-& \frac{n_1 k_B T}{q} \,
          W\Big\{\frac{q}{n_1 k_B T}\,I_{01}R_{p1} \,
               \mathrm{exp}\Big[
                  \frac{q}{n_1 k_B T}\,R_{p1}\,(I + I_{ph} + I_{01})
                    \Big]
             \Big\}
\end{eqnarray}
This formula can be re-written just as for the one diode
model.  
We obtain
\begin{equation}
\label{eqx1}
  x_1 = \mathrm{log}\Big[\frac{q}{n_1 k_B T}\,I_{01}R_{p1}
                \Big]
       + \frac{q}{n_1 k_B T}\,R_{p1}\,(I + I_{ph} + I_{01}).
\end{equation}
and
\begin{equation}
\label{eqv1}
  \frac{q}{n_1 k_B T}\,V_1 
         + \mathrm{log}\Big[
                              \frac{q}{n_1 k_B T}\,I_{01}R_{p1}
                           \Big] = g(x_1).
\end{equation}

Letting $V_2$ equal the voltage drop across subcircuit 2,
which includes the reverse bias diode and shunt resistance,
equation (9) of \cite{RDPA-2012} is a comparable
situation, with no photocurrent source.
That is,
\begin{eqnarray}
  \nonumber
  V_2 &=& f_2(I) = (I - I_{02}) \, R_{p2} \\
        \label{eqrdpa9}
        &+& \frac{n_2 k_B T}{q} \,
          W\Big\{\frac{q}{n_2 k_B T}\,I_{02}R_{p2} \,
               \mathrm{exp}\Big[
                  \frac{-q}{n_2 k_B T}\,R_{p2}\,(I - I_{02})
                    \Big]
             \Big\}
\end{eqnarray}
That formula also can be rewritten.
We let
\begin{equation}
\label{eqx2}
  x_2 = \mathrm{log}\Big[\frac{q}{n_2 k_B T}\,I_{02}R_{p2}
                \Big]
       - \frac{q}{n_2 k_B T}\,R_{p2}\,(I - I_{02}).
\end{equation}
The result is
\begin{equation}
\label{eqv2}
  - \frac{q}{n_2 k_B T}\,V_2
         + \mathrm{log}\Big[
                              \frac{q}{n_2 k_B T}\,I_{02}R_{p2}
                           \Big] = g(x_2).
\end{equation}
Once again we see that equations 
(\ref{eqx2}) and (\ref{eqv2}) represent the 
relationship between $I$ and $V_2$ in subcircuit 2
as a shift of origin and scale change of the current $I$,
related to a shift of origin and scale change
of the voltage $V_2$, the relationship being
given by the function $y = g(x)$.
The origin shifts and scale changes are however
different from those for subcircuit 1, 
both for the current transformations
and for the voltage transformations.

The two origin shifts and scale changes for the
current are given by equations (\ref{eqx1}) and (\ref{eqx2}).
The two origin shifts and scale changes for the
voltage are given by equations (\ref{eqv1}) and (\ref{eqv2}).

The voltage $V$ across the solar cell is given by
equation (1) of \cite{RDPA-2012},
\begin{equation}
\label{eqrpda1}
  V = I\,R_s + V_1 + V_2
\end{equation}
In terms of the function $g(x)$, using the 
two variables $x_1$ and $x_2$ which are
transformations of the current $I$, the
voltage $V$ is calculated as
\begin{eqnarray}
  \nonumber
  V = I\,R_s &+& n_1 \, \frac{k_B T}{q}\,g(x_1)
                  -  n_2 \, \frac{k_B T}{q}\,g(x_2)\\
                  \nonumber
                  &-&  \frac{n_1 k_B T}{q} \mathrm{log}\Big[
                              \frac{q}{n_1 k_B T}\,I_{01}R_{p1}
                           \Big]\\
                  \label{eqvtot}
                  &+& \frac{n_2 k_B T}{q} \mathrm{log}\Big[
                              \frac{q}{n_2 k_B T}\,I_{02}R_{p2}
                           \Big].
\end{eqnarray}
Equation (\ref{eqvtot}), with the ancillary equations
(\ref{eqx1}) and (\ref{eqx2}) to define $x_1$ and $x_2$,
is our proposed replacement for formula (\ref{eqrdpa}).

The new formula (\ref{eqvtot}) produces the same 
$I-V$ characteristic curve as the original formula
(\ref{eqrdpa}).  
The parameters above, used to draw figure \ref{fig-rdpa2},
were obtained by the authors of \cite{RDPA-2012}
from a best fit of the two diode model to an
actual solar cell, 
with the series resistance $R_s$ forced to zero.
With these parameter values,
in the original formula there would be arithmetic
overflow even if hardware quadruple precision arithmetic
were used, since
the argument to one of the Lambert W function 
evaluations is about $10^{11232}$
when using the formula (\ref{eqrdpa}) at $I=0$.
In contrast, for the new formula (\ref{eqvtot}), 
there is no overflow
since the maximum magnitude of an intermediate
variable involved in the $g(x)$ function calculation
is less than 30000, and the maximum magnitude of
an intermediate variable involved in the overall $V = f(I)$
calculation of (\ref{eqvtot}) is about $10^8$.

We can visualize the formula (\ref{eqvtot}) geometrically
as the sum of three curves, a sloped straight line plus two 
flipped copies of the $y = g(x)$ curve.
One copy of the $y = g(x)$ curve, flipped, shifted,
and scaled proportionally to the ideality factor $n_1$,
is to be added; it represents the subcircuit 1 voltage 
$V_1$, the concave-up
portion of the $I-V$ curve.
The second copy of the $y = g(x)$ curve, flipped, shifted,
and scaled proportionally to the ideality factor $n_2$,
is to be subtracted; it represents the 
subcircuit 2 voltage $V_2$, the concave-down portion of
the $I-V$ curve.

The authors of \cite{GSLMMOC-2013} have also proposed
a three diode model which provides an improved fit for
an organic solar cell with two inflection points in its
$I-V$ characteristic curve.  
Their model presents some interesting computational
challenges, related to assumptions regarding the
ideality factors of the various diodes in the model.
We have not considered, in this present paper,
the applicability of our calculational methodology to
the three diode model of \cite{GSLMMOC-2013}.
We believe that represents an opportunity for further
exploration.

\section{Implementing the $y = g(x)$ Function}
\label{sect-galgo}

In this section, we describe the implementation of
the calculation of the $y = g(x)$ function.
See \cite{Roberts-2015} for full details,
including a discussion of the computational stability
considerations.
Here we present only a straightforward description
of the necessary calculations to obtain $g(x)$.

Context:  The function 
$y = g(x) = \mathrm{log}(W(\mathrm{exp}(x)))$
is to be calculated.
The variable $x$ is a real number argument,
positive or negative.
The result variable $y$ is also a real number.
See figures \ref{fig-logwexp1} and \ref{fig-logwexp2}
for the graph of $y = g(x)$.
The computer language should have available exponential
and logarithm functions for the chosen precision,
and a stored value of $e = \mathrm{exp}(1.0)$.

(a) Make an initial estimate $y_0$ of the result, 
as follows.

For $x \le -e$, take $y_0 = x$.

For $x \ge e$, take $y_0 = \mathrm{log}(x)$.

For $-e < x < e$, take $y_0$ as a linear interpolation
between the points $(-e,-e)$ and $(e,1)$.
That is, 
\begin{equation*}
 y_0 = -e + \frac{1+e}{2e} \, (x+e)
\end{equation*}

\begin{figure}
\makebox[\textwidth][c]{
  \includegraphics[scale=0.60]{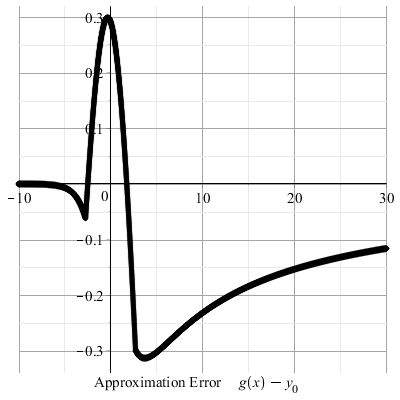}}
{
\captionsetup{format=hang}
  \caption{
\label{fig-y0err}Graph of $g(x) - y_0$ for approximation error\\
  for initial estimate of g(x).\\
  ---------------------------------------------------------------------
}}
\end{figure}

This estimate $y_0$ is very crude.  
See figure \ref{fig-y0err} for a graph of the absolute error
$g(x) - y_0$ for $-10 < x < 30$.
The absolute error lies between -0.32 and +0.30.

(b) Refine the estimate $y_0$ by calculating
\begin{equation*}
  y_1 = y_0 - 
  \frac{2 (y_0 + e^{y_0} - x) (1 + e^{y_0})}
         {2 (1 + e^{y_0})^2 - (y_0 + e^{y_0} - x) e^{y_0}}.
\end{equation*}

\begin{figure}
\makebox[\textwidth][c]{
  \includegraphics[scale=0.60]{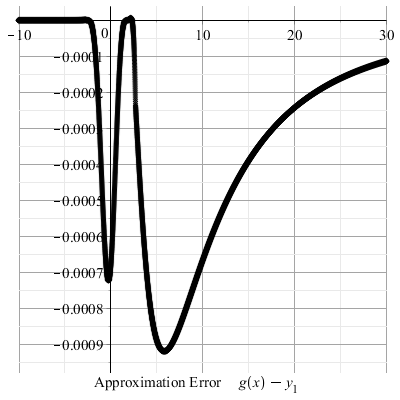}}
{
\captionsetup{format=hang}
  \caption{
\label{fig-y1err}Graph of $g(x) - y_1$ for approximation error\\
  for first iterative refinement of estimate of g(x).\\
  ---------------------------------------------------------------------
}}
\end{figure}

This iteration formula is Halley's method and has cubic convergence.
See figure \ref{fig-y1err} for a graph of the absolute error
$g(x) - y_1$ for $-10 < x < 30$.
The absolute error lies between -0.001 and +0.00001.

(c) Further refine the estimate $y_1$ by calculating
\begin{equation*}
  y_2 = y_1 - 
  \frac{2 (y_1 + e^{y_1} - x) (1 + e^{y_1})}
         {2 (1 + e^{y_1})^2 - (y_1 + e^{y_1} - x) e^{y_1}}.
\end{equation*}

\begin{figure}
\makebox[\textwidth][c]{
  \includegraphics[scale=0.60]{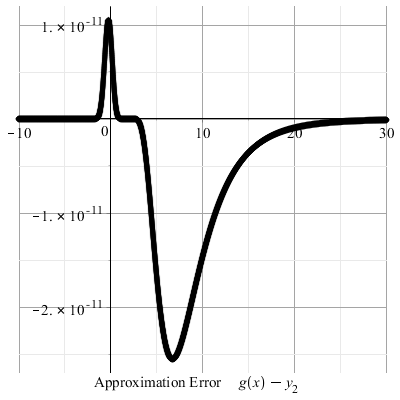}}
{
\captionsetup{format=hang}
  \caption{
\label{fig-y2err}Graph of $g(x) - y_2$ for approximation error\\
  for second iterative refinement of estimate of g(x).\\
  ---------------------------------------------------------------------
}}
\end{figure}

This second iteration produces an estimate which is good
enough for most practical purposes.
See figure \ref{fig-y2err} for a graph of the absolute error
$g(x) - y_2$ for $-10 < x < 30$.
The absolute error lies between $-10^{-10}$ and $+10^{-10}$.

The computational workload to obtain $y_2$ is two
evaluations of the exponential function and one
evaluation of the logarithm function.

(d) If the estimate $y_2$ is not precise enough
for the intended application, perform additional
iterations of the refinement formula,
\begin{equation*}
  y_{n+1} = y_n - 
  \frac{2 (y_n + e^{y_n} - x) (1 + e^{y_n})}
         {2 (1 + e^{y_n})^2 - (y_n + e^{y_n} - x) e^{y_n}}.
\end{equation*}

\begin{figure}
\makebox[\textwidth][c]{
  \includegraphics[scale=0.60]{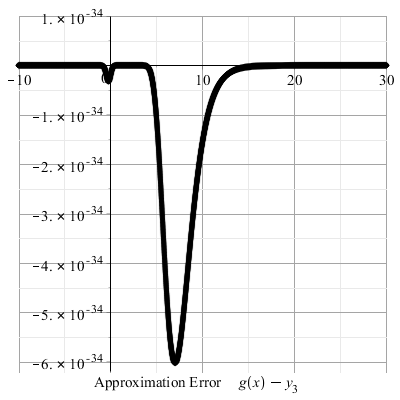}}
{
\captionsetup{format=hang}
  \caption{
\label{fig-y3err}Graph of $g(x) - y_3$ for approximation error\\
  for third iterative refinement of estimate of g(x).\\
  ---------------------------------------------------------------------
}}
\end{figure}

See figure \ref{fig-y3err} for a graph of the absolute error
$g(x) - y_3$ after three iterations, for $-10 < x < 30$.
The absolute error lies between $-10^{-33}$ and $+10^{-55}$.

Note: If examining the differences $|y_{n+1} - y_n|$
to determine if the result has converged, one should
test for the absolute error, not the relative error.
Because $y = 0$ for $x = 1$, a relative error test
would be unreliable for $x$ in the vicinity of 1.
Or, one might wish to code to test for absolute
error if $|y_n|$ were less than 1, and for relative
error for larger magnitudes of $|y_n|$.

The reason the above algorithm does not produce arithmetic
overflow is that at no point is $\mathrm{exp}(x)$ calculated.
Rather, the calculations involve only $x$, $y$,
and $\mathrm{exp}(y)$.
If hardware arithmetic underflow produces an error exception,
instead of a zero result, then it will be desirable to code a
test for negative values of $y_n$, for example $y_n < -100$,
and simply stop iterating in that circumstance.
Further discussion of the method is in reference \cite{Roberts-2015}.

\section{Conclusion}

We have suggested alternative formulas for computing
the explicit function $V = f(I)$ which gives the exact
current-voltage characteristic of the one or two diode 
models of a solar cell,
in particular a solar cell with the S-curve property.
The alternative formulas are less likely to produce 
arithmetic overflow errors when calculated with
hardware floating point arithmetic.
The alternative formulas are suitable for implementation
in Fortran or C, or on micro-controllers.

We thank B. Romero and F. J. Garc\'ia-S\'anchez for their
helpful comments during our investigations.


\end{document}